\documentclass[12pt,a4paper]{article}
\usepackage{amssymb}
\usepackage[dvips]{graphicx}
\usepackage{psfrag}

\newcommand{\sss}{\scriptscriptstyle}
\renewcommand{\Re}{{\rm Re}}

\begin{document}
\begin{flushright}
UAB--FT--634\\
November 2007
\end{flushright}
\vspace*{0.6cm}

\begin{center}
{\Large\bf
Sum rules for {\boldmath $B\to \pi\eta^{(\prime)}, K\eta^{(\prime)}, \eta^{(\prime)}\eta^{(\prime)}$}  decays}
\vspace*{1cm}

Rafel Escribano, Joaquim Matias and Javier Virto
\vspace*{0.2cm}

{\footnotesize\it
Grup de F\'{\i}sica Te\`orica and IFAE, Universitat Aut\`onoma de Barcelona,\\
E-08193 Bellaterra (Barcelona), Spain}

\end{center}
\vspace*{0.2cm}

\begin{abstract}
We provide a set of sum rules, using flavour symmetries,
relating CP-averaged ratios and CP asymmetries of different neutral and charged $B$ mesons
decaying into an $\eta^{(\prime)}$ particle together with a pion, a kaon or a second $\eta^{(\prime)}$.
These sum rules allow us to give a prediction for the $B^0 \to K^0 \eta$ branching ratio.
We also predict a clear sign anti-correlation between the two
$B^0 \to \pi^0 (\eta,\eta^\prime)$ CP asymmetries,
and find a  combined constraint on the branching ratios and CP asymmetries of the three
$B \to\eta^{(\prime)} \eta^{(\prime)}$ decay modes.
\end{abstract}

\section{Introduction}
\label{intro}

$B$-physics is entering a golden epoch due to the huge amount of available data on $B_d$ decays coming from the $B$-factories \cite{babar,belle},
the interesting $B_s$ decay channels measured at Tevatron \cite{cdf} and, in the near future, the plethora of decays that will be observed at
LHC \cite{lhcb}.
They provide  many different strategies of testing the Standard Model (SM) and looking for possible ``smoking gun'' signals coming from New Physics (NP).
One of such strategies consists in constructing observables, based on certain $B$ decay channels
($B\to K^* l^+l^-$, $B\to \pi K$, $B\to \pi\pi$, $B \to KK$, ...),
that can test the presence of specific types of NP, for instance, observables sensitive to the presence of right-handed currents \cite{rh},
isospin breaking induced by NP (see for example Ref.~\cite{su3fv1}), etc.

Sum rules have been used as a way of constructing observables sensitive to isospin or $SU(3)$ breaking.
In order to extract useful information from this type of sum rules it is necessary to be able to estimate the expected size of the isospin or
$SU(3)$ breaking.
Different approaches  exist in the literature that may help in principle to estimate the size of this type of breaking: flavor symmetries
\cite{su3fv2,fs,Chiang:2006ih}, QCD factorization techniques \cite{martin,jm},
SCET \cite{scet}, or a combination of flavor symmetries with QCD factorization \cite{dmv}.

However, these type of sum rules admit a twofold reading depending on the availability of data.
On the one side, if all observables entering the sum rule are known, then the sum rule can serve as a test of the size of
the $SU(3)$ breaking.
If the same parameter enters another sum rule, we automatically gain control on the size of the $SU(3)$ breaking in the later sum rule.
Moreover, if the obtained $SU(3)$-breaking parameter points towards a too large breaking it could be a signal of the presence of isospin or
$SU(3)$-breaking NP contributions.
An example of this type of analysis is the Lipkin sum rule \cite{jm,lipkin,ne} of the recent $B \to\pi K$ puzzle \cite{puzzle}.
On the other side, given that the sum rule is a combination of observables
(usually branching ratios and CP asymmetries)
that should sum up to zero up to the estimated isospin or $SU(3)$-breaking contributions,
they allow to extract information on the not yet measured observables entering the sum rule.

In this Letter we present new sum rules involving $B\to \pi\eta^{(\prime)}$, $K\eta^{(\prime)}$
and $\eta^{(\prime)}\eta^{(\prime)}$ decays.
One of them will be a function of measured observables and will serve us as a test of the size of the
$SU(3)$ breaking.
The rest of sum rules will provide relations between observables including not yet measured branching ratios and
CP asymmetries and they will allow us to obtain some predictions.
The present work is an extension of a previous paper \cite{jm}, where one of us (JM) studied a series of sum rules for $B\to \pi K$
decays in the framework of QCD factorization.
Here we extend those ideas to include $B\to \pi\eta^{(\prime)}$, $K\eta^{(\prime)}$ and
$\eta^{(\prime)}\eta^{(\prime)}$ decays with two important differences.
First, in Ref.~\cite{jm} isospin breaking referred to the contributions of all weak operators with
$\Delta I\neq 0$.
Their contributions were written in terms of suppressed ratios of the type `$T/P$'
(tree versus penguin amplitudes).
In this Letter, we will have isospin and also $SU(3)$-breaking contributions that will include
any contribution sensitive to mass differences between up,
down and strange quarks ($\eta$-$\eta^\prime$ mixing, etc.) \cite{su3fv2}.
We will explicitly distinguish between $SU(3)$-breaking effects induced by
$\eta$-$\eta^\prime$ mixing, unavoidable when dealing with $\eta$ or $\eta^\prime$
mesons in the final state, and other possible sources of $SU(3)$ breaking.
Second, the way to deal with the subleading contributions of the type `$T/P$' is different.
One approach is to evaluate them explicitly using QCD factorization,
as it was done in Ref.~\cite{jm} for $B\to \pi K$ sum rules.
A different strategy may be to try to find a hierarchy between the different subleading terms \cite{su3fv2}.
Finally, a third possibility is to relate those subleading terms to $\Delta S=0$ processes using flavor symmetries
(similar to what was done in Ref.~\cite{flav} when relating the $B_s \to K^+K^-$ decay with $B\to \pi\pi$ using $U$-spin \cite{u-spin}).
In this Letter we will follow this last approach.
This means in practice that those subleading terms are moved from the r.h.s of the sum rule to the
l.h.s and they are expressed in terms of observables.

The outline of this paper is the following.
In Section 2, we discuss the $SU(3)$ decomposition of amplitudes and define the observables that will enter the sum rules.
In Section 3, we present a set of six new sum rules and discuss them in turn.
We focus on the information that can be extracted for the branching ratio of $B\to K^0\eta$,
the CP asymmetries of $B^0 \to \pi^0\eta$ and $B^0 \to \pi^0\eta^\prime$,
and the three neutral decays $B^0 \to \eta\eta$, $\eta\eta^\prime$ and $\eta^\prime\eta^\prime$.
We also pin down the main source of error affecting the different sum rules and study the impact that reducing those errors would
have on our observables.
We conclude in Section 4.

\section{Amplitudes and observables}
\label{amplitudes}

The decay amplitudes of the modes under consideration can be written in terms of the basis of
$T$ (tree), $P$ (penguin), $C$ (color-suppressed tree), $E$ (exchange), $A$ (annihilation), and
$PA$ (penguin annihilation) diagram contributions \cite{Gronau:1994rj,Dighe:1995bm}.
The contributions $E$, $A$, and $PA$ are usually neglected since they are formally
suppressed by a factor of $f_B/m_B=5\%$ \cite{Gronau:1995hn}.
$E$ and $A$ are also helicity suppressed by a factor $m_q/m_b$ where $q=u,d,s$.
However, they may be enhanced through rescattering effects (see Ref.~\cite{res}).
These rescattering effects could be tested by comparing the $\Delta C=\Delta S=0$ transitions
$B_s \to  \pi^+ K^-,\pi^0 {\bar K^0}$ and $\eta_8 {\bar K^0}$,
which are unaffected by those topologies,
with the partners transitions $B \to \pi^+ \pi^-, \pi^0 \pi^0, \pi^0 \eta_8$, and $\eta_8 \eta_8$
which receive $E$ and $PA$ contributions.
In the diagrammatic basis, the set of required amplitudes are written in terms of three independent combinations, the so-called $t$, $p$, and $c$ for $\Delta S=0$ transitions and
$t^\prime$, $p^\prime$, and $c^\prime$ for $|\Delta S|=1$.
In the approximation of neglecting the $E$, $A$, and $PA$ contributions,
the former combinations are identified as \cite{Chiang:2006ih}
\begin{equation}
\label{tctcprime}
\begin{array}{l}
t\equiv Y^u_{db}\,T-\left(Y^u_{db}+Y^c_{db}\right)P_{EW}^{C}\ , \quad
t^\prime\equiv Y^u_{sb}\,\xi_T\,T-\left(Y^u_{sb}+Y^c_{sb}\right)P_{EW}^{C}\ ,\\[2ex]
c\equiv Y^u_{db}\,C-\left(Y^u_{db}+Y^c_{db}\right)P_{EW}\ , \quad
c^\prime\equiv Y^u_{sb}\,\xi_C\,C-\left(Y^u_{sb}+Y^c_{sb}\right)P_{EW}\ ,
\end{array}
\end{equation}
for the tree amplitudes, where $P_{EW}$ and $P^C_{EW}$ are color-favored and color-suppressed
electroweak penguin amplitudes, respectively, and
\begin{equation}
\label{pprime}
\begin{array}{l}
p\equiv -\left(Y^u_{db}+Y^c_{db}\right)\left(P-\frac{1}{3}P_{EW}^{C}\right)\ ,\\[2ex]
p^\prime\equiv
-\left(Y^u_{sb}+Y^c_{sb}\right)\left(\xi_P\,P-\frac{1}{3}P_{EW}^{C}\right)\
,
\end{array}
\end{equation}
for the corresponding penguin amplitudes.
In these expressions,
$Y^{q^\prime}_{qb}\equiv V_{q^\prime q}V^\ast_{q^\prime b}$ ($q^\prime$ can be either $u$ or $c$)
and $\xi_T$, $\xi_C$ and $\xi_P$ are $SU(3)$-breaking factors for the $T$, $C$ and $P$
amplitudes, respectively, when going from $\Delta S=0$ to $|\Delta S|=1$ transitions.
In addition to the former octet-type contributions, there are also singlet-type contributions that
must be incorporated when the pseudoscalar final state contains $\eta$ and/or $\eta^\prime$ mesons.
In the diagrammatic approach these singlet contributions are expressed in terms of $t_s$, $p_s$
(usually $s=p_s/3$ is introduced instead of $p_s$), $c_s$, and $s_0$ ($s_0$ contributes only to
$\eta_0\eta_0$ decays), where
\begin{equation}
\label{ssprime}
\begin{array}{l}
s\equiv -\left(Y^u_{db}+Y^c_{db}\right)\left(S-\frac{1}{3}P_{EW}\right)\ ,\\[2ex]
s^\prime\equiv -\left(Y^u_{sb}+Y^c_{sb}\right)\left(\xi_S\,S-\frac{1}{3}P_{EW}\right)\ ,
\end{array}
\end{equation}
with $S$ the singlet penguin contribution and $\xi_S$ the $SU(3)$-breaking factor.
The $t_s$, $c_s$, and $s_0$ will be neglected when obtaining the sum rules\footnote{
There is certain controversy concerning the size of the $t_s$, $c_s$ and
$s_0$ contributions.
While the usual assumption \cite{rosner1} is to neglect those terms and keep $p_s$, as we
do here and seems to be allowed by experiment, in SCET,
assuming certain scaling of the operators, those terms may play some role~\cite{zupan}.
However, experimental data should first confirm if they are needed or not.}.
So the only significant additional contributions one has to include are the $s$-type contributions.

\begin{table}
\centerline{
{\small
\begin{tabular}{lc}
\hline\hline
Mode & Amplitude\\
\hline\\[-1ex]
$B^+\to K^+\eta$ & $\frac{1}{\sqrt{6}}\left[
(c_\theta-\sqrt{2}s_\theta)(t+c)-(c_\theta+2\sqrt{2}s_\theta)p-3\sqrt{2}s_\theta s\right]$\\[1ex]
$B^0\to K^0\eta$ & $\frac{1}{\sqrt{6}}\left[
(c_\theta-\sqrt{2}s_\theta)c-(c_\theta+2\sqrt{2}s_\theta)p-3\sqrt{2}s_\theta s\right]$\\[1ex]
$B^+\to K^+\eta^\prime$ & $\frac{1}{\sqrt{6}}\left[
(\sqrt{2}c_\theta+s_\theta)(t+c)+(2\sqrt{2}c_\theta-s_\theta)p+3\sqrt{2}c_\theta s\right]$\\[1ex]
$B^0\to K^0\eta^\prime$ & $\frac{1}{\sqrt{6}}\left[
(\sqrt{2}c_\theta+s_\theta)c+(2\sqrt{2}c_\theta-s_\theta)p+3\sqrt{2}c_\theta s\right]$\\[1ex]
\hline\\[-2ex]
$B^+\to \pi^+\eta$ & $\frac{1}{\sqrt{6}}\left[
(c_\theta-\sqrt{2}s_\theta)(2p+t+c)-3\sqrt{2}s_\theta s\right]$\\[1ex]
$B^0\to \pi^0\eta$ & $\frac{1}{2\sqrt{3}}\left[
2(c_\theta-\sqrt{2}s_\theta)p-3\sqrt{2}s_\theta s\right]$\\[1ex]
$B^+\to \pi^+\eta^\prime$ & $\frac{1}{\sqrt{6}}\left[
(\sqrt{2}c_\theta+s_\theta)(2p+t+c)+3\sqrt{2}c_\theta s\right]$\\[1ex]
$B^0\to \pi^0\eta^\prime$ & $\frac{1}{2\sqrt{3}}\left[
2(\sqrt{2}c_\theta+s_\theta)p+3\sqrt{2}c_\theta s\right]$\\[1ex]
$B^0\to \eta\eta$ & $\frac{1}{3}\left[
(1 + s_\theta(s_\theta-2\sqrt{2}c_\theta))(p+c)-3\sqrt{2}s_\theta(c_\theta-\sqrt{2}s_\theta) s\right]$\\[1ex]
$B^0\to \eta^\prime\eta^\prime$ & $\frac{1}{3}\left[
(1+c_\theta(c_\theta+2\sqrt{2}s_\theta))(p+c)+3\sqrt{2}c_\theta(s_\theta+\sqrt{2}c_\theta) s\right]$\\[1ex]$B^0\to \eta\eta^\prime$ & $\frac{1}{6}\left[
(2\sqrt{2}c_{2\theta}-s_{2\theta})(p+c)+3\sqrt{2}(c_{2\theta}-\sqrt{2}s_{2\theta}) s\right]$\\[1ex]
\hline\hline
\end{tabular}
}}
\caption{\small
Diagrammatic decomposition of the $|\Delta S|=1$ \emph{(upper part)} and
$\Delta S=0$ \emph{(bottom part)} transitions of $B$ decays involving
$\eta$ and $\eta^\prime$ mesons.}
\label{tableetaetaprime}
\end{table}

The diagrammatic decomposition of the relevant $|\Delta S|=1$ and $\Delta S=0$ transitions involving $\eta$ and $\eta^\prime$ mesons is shown in
Table \ref{tableetaetaprime}. From this table, the following amplitude relation can be written\footnote{ For the members of the pseudoscalar meson
nonet and the triplet of $B$'s, we use the convention of Ref.~\cite{Grinstein:1996us}. The physical mesons $\eta$ and $\eta^\prime$ are defined
through the rotation
\[
\left(
\begin{array}{c}
\eta\\[1ex]
\eta^\prime
\end{array}
\right)=
\left(
\begin{array}{cc}
-\cos\theta & +\sin\theta\\[1ex]
-\sin\theta & -\cos\theta
\end{array}
\right)
\left(\begin{array}{c}
\eta_8\\[1ex]
\eta_0
\end{array}
\right)\ ,
\]
where the sign convention is such that the angle $\theta$ agrees with the PDG \cite{Yao:2006px}.
The current experimental value for the mixing angle is
$\theta=(-13.3\pm 1.0)^\circ$ \cite{Ambrosino:2006gk}. }:
\begin{equation}
(\sqrt{2}c_\theta+s_\theta)(A(K^+\eta)-A(K^0\eta))=
(c_\theta-\sqrt{2}s_\theta)(A(K^+\eta^\prime)-A(K^0\eta^\prime))\ ,
\end{equation}
where $\eta$-$\eta^\prime$ mixing is admitted as the only source of $SU(3)$ breaking.
Notice that this relation is deduced only after neglecting the $t_s$ contribution,
which only affects the $B^+\to K^+\eta_0$ transition.
Other interesting relations are
\begin{equation}
\begin{array}{c}
(c_\theta-\sqrt{2}s_\theta)(A(K^+\pi^-)+A(K^0\pi^+))-\sqrt{6}(A(K^+\eta)-A(K^0\eta))=0\ ,\\[1ex]
(s_\theta+\sqrt{2}c_\theta)(A(K^+\pi^-)+A(K^0\pi^+))-\sqrt{6}(A(K^+\eta^\prime)-A(K^0\eta^\prime))=0\ ,
\end{array}
\end{equation}
where a combination of them can already be found in Ref.~\cite{su3fv2}.

The collected experimental data on the branching ratios of the studied decay modes are organized in two types of observables.
A first type of observables are the ratios of CP-averaged branching ratios $R_c$, $R_0$ and $R$,
defined in Refs.~\cite{Fleischer:1997um,Buras:1998rb}, and
$R_c^{\pi\pi}=BR(B^+\to \pi^+\pi^0)/BR(B^+\to K^0\pi^+)$.
We also define
\begin{equation}
\label{avgBReta}
\begin{array}{c}
R_c^{K\eta}=\frac{\displaystyle \mbox{BR}(B^+\to K^+\eta)}{\displaystyle \mbox{BR}(B^+\to K^0\pi^+)}\ ,\quad
R_0^{K\eta}=\frac{\displaystyle \mbox{BR}(B^0\to K^0\eta)}{\displaystyle \mbox{BR}(B^+\to K^0\pi^+)}\ ,
\\[2ex]
R_0^{\pi\eta}=\frac{\displaystyle \mbox{BR}(B^0\to \pi^0\eta)}{\displaystyle \mbox{BR}(B^0\to K^0\pi^+)}\ ,
\\[2ex]
\end{array}
\end{equation}
and the same for the associated $\eta^\prime$ channels.
The decay $B^+\to K^0\pi^+$, governed by the penguin amplitude $p'$, is used as the normalization process for all ratios.
CP asymmetries are the second type of observables.
As in Ref.~\cite{jm}, we denote by
${\cal A}_{\rm CP}^{+0}$, ${\cal A}_{\rm CP}^{00}$, ${\cal A}_{\rm CP}^{+-}$, and
${\cal A}_{\rm CP}^{0+}$,
the CP asymmetries\footnote{
We follow the PDG notation \cite{Yao:2006px}, which agrees with that of the HFAG \cite{HFAG},
for the definitions of the CP-averaged decay widths and the direct CP asymmetries.
Notice, however, the different sign notation of the asymmetries with respect to Ref.~\cite{jm}.}
of the $K^+\pi^0$, $K^0\pi^0$, $K^+\pi^-$, and $K^0\pi^+$ channels, respectively.
We also define
\begin{equation}
\label{CPasym}
\begin{array}{rcl}
{\cal A}_{\rm CP}^{\pi^+\pi^0} &=&
\frac{\displaystyle \Gamma(B^-\to \pi^-\pi^0)-\Gamma(B^+\to \pi^+\pi^0)}
        {\displaystyle \Gamma(B^-\to \pi^-\pi^0)+\Gamma(B^+\to \pi^+\pi^0)}\ ,\\[2ex]
{\cal A}_{\rm CP}^{K^0\bar K^0} &=&
\frac{\displaystyle \Gamma(\bar B^0\to K^0\bar K^0)-\Gamma(B^0\to K^0\bar K^0)}
        {\displaystyle \Gamma(\bar B^0\to K^0\bar K^0)+\Gamma(B^0\to K^0\bar K^0)}\ ,
\end{array}
\end{equation}
together with the new definitions (also for the $\eta^\prime$ channels)
\begin{equation}
\label{CPasymeta}
\begin{array}{rcl}
{\cal A}_{\rm CP}^{K^+\eta} &=&
\frac{\displaystyle \Gamma(B^-\to K^-\eta)-\Gamma(B^+\to K^+\eta)}
        {\displaystyle \Gamma(B^-\to K^-\eta)+\Gamma(B^+\to K^+\eta)}\ ,\\[2ex]
{\cal A}_{\rm CP}^{K^0\eta} &=&
\frac{\displaystyle \Gamma(\bar B^0\to \bar{K}^0\eta)-\Gamma(B^0\to K^0\eta)}
        {\displaystyle \Gamma(\bar B^0\to \bar{K}^0\eta)+\Gamma(B^0\to K^0\eta)}\ ,\\[2ex]
{\cal A}_{\rm CP}^{\pi^0\eta} &=&
\frac{\displaystyle \Gamma(\bar B^0\to \pi^0\eta)-\Gamma(B^0\to \pi^0\eta)}
        {\displaystyle \Gamma(\bar B^0\to \pi^0\eta)+\Gamma(B^0\to \pi^0\eta)}\ ,
\end{array}
\end{equation}
and
\begin{equation}
\label{CPasymetaeta}
\begin{array}{rcl}
{\cal A}_{\rm CP}^{\eta\eta} &=&
\frac{\displaystyle \Gamma(\bar B^0\to \eta\eta)-\Gamma(B^0\to \eta\eta)}
        {\displaystyle \Gamma(\bar B^0\to \eta\eta)+\Gamma(B^0\to \eta\eta)}\ ,\\[2ex]
{\cal A}_{\rm CP}^{\eta^\prime\eta^\prime} &=&
\frac{\displaystyle \Gamma(\bar B^0\to \eta^\prime\eta^\prime)-\Gamma(B^0\to \eta^\prime\eta^\prime)}
        {\displaystyle \Gamma(\bar B^0\to \eta^\prime\eta^\prime)+\Gamma(B^0\to \eta^\prime\eta^\prime)}\ ,\\[2ex]
{\cal A}_{\rm CP}^{\eta\eta^\prime} &=&
\frac{\displaystyle \Gamma(\bar B^0\to \eta\eta^\prime)-\Gamma(B^0\to \eta\eta^\prime)}
        {\displaystyle \Gamma(\bar B^0\to \eta\eta^\prime)+\Gamma(B^0\to \eta\eta^\prime)}\ .
\end{array}
\end{equation}

The main purpose of this work is to provide a set of model independent sum rules relating CP-averaged
branching ratios and CP asymmetries of different non-leptonic $B\to hh^\prime$ decays with
$h,h^\prime=\pi,K,\eta,\eta^\prime$.
The list of measured observables to be used in our analysis is shown in Table \ref{tableexpvalues}.
In order to be able to write exact sum rules for $|\Delta S|=1$ processes including $\eta$ and $\eta^\prime$,
information coming from  $\Delta S=0$ decay modes is required.

\begin{table}
\centerline{
\begin{tabular}{lcccc}
\hline\hline
Mode & $BR_{\rm exp}$ & ${\cal A}_{\rm CP}^{\rm exp}$ & $R_{\rm exp}$                                   & Refs.\\[0.5ex]
\hline
$B^+\to K^+\pi^0$ & $12.8\pm 0.6$ & $0.047\pm 0.026$ & $1.11\pm 0.07$                                   & \cite{exp1}\\
$B^0\to K^0\pi^0$ & $10.0\pm 0.6$ & $-0.12\pm 0.11$ & $0.87\pm 0.06$                                    & \cite{exp1,exp2}\\
$B^0\to K^+\pi^-$  & $19.4\pm 0.6$ & $-0.095\pm 0.013$ & $0.84\pm 0.04$                                 & \cite{exp1,exp3,exp3bis}\\
$B^0\to K^0\pi^+$ & $23.1\pm 1.0$ & $0.009\pm 0.025$ & ---                                              & \cite{exp1,exp4}\\
$B^+\to K^+\eta$   & $2.2\pm 0.3$   & $-0.29\pm 0.11$ & $0.10\pm 0.01$                                  & \cite{exp5,exp6}\\
$B^+\to K^+\eta^\prime$ & $69.7^{+2.8}_{-2.7}$ & $0.031\pm 0.021$ & $3.02\pm 0.18$                      & \cite{exp7,exp8,exp9}\\
$B^0\to K^0\eta^\prime$ & $64.9\pm 3.5$ & $0.09\pm 0.06$ & $2.81\pm 0.19$                               & \cite{exp7,exp8,exp9}\\
$B^+\to \pi^+\pi^0$ & $5.7\pm 0.4$ & $0.04\pm 0.05$ & $0.25\pm 0.02$                                    & \cite{exp1}\\
$B^+\to K^0\bar K^0$ & $0.96^{+0.21}_{-0.19}$ & $-0.58^{+0.73}_{-0.66}$ &  ---                          & \cite{exp4,exp10}\\
$B^0\to \pi^0\eta$ & $0.6^{+0.5}_{-0.4}$ & --- & $0.026\pm 0.020$                                       & \cite{exp6,exp11}\\
$B^0\to \pi^0\eta^\prime$ & $1.5^{+0.7}_{-0.6}$ & --- & $0.065\pm 0.030$                                & \cite{exp8,exp11}\\
$B^0\to \eta\eta$ & $1.1^{+0.5}_{-0.4}$ & --- & ---                                                     & \cite{exp6,exp12}\\
$B^0\to \eta^\prime\eta^\prime$ & $1.0^{+0.8}_{-0.6}$ & --- & ---                                       & \cite{exp12}\\
$B^0\to \eta\eta^\prime$ & $0.2^{+0.8}_{-0.6}$ & --- & ---                                              & \cite{exp11}\\
\hline
\end{tabular}
}
\caption{\small
Experimental values \cite{HFAG} of the observables that enter the sum rules.}
\label{tableexpvalues}
\end{table}

\section{Sum rules}
\label{sumrules}

In this section we present six new sum rules relating $B \to K\eta^{(\prime)}$, $B \to \pi \eta^{(\prime)}$
and $B^0 \to \eta^{(\prime)} \eta^{(\prime)}$ branching ratios and CP asymmetries.
The first sum rule (I) serves as a test of the control we have on the size of the $SU(3)$-breaking effects.
In the second sum rule (II) we obtain a correlation between the
CP asymmetries ${\cal A}_{\rm CP}^{\pi^0 \eta}$ and ${\cal A}_{\rm CP}^{\pi^0 \eta^\prime}$.
The next sum rule (III) allows to predict the CP-averaged branching ratio of the decay $B^0 \to K^0 \eta$,
which combined with the fourth sum rule (IV) can, in principle, provide a value for  the direct CP asymmetry of $B^0 \to K^0 \eta$.
The last two sum rules (V and VI) exhibit relations between $\Delta S=0$ observables alone,
that can be useful to obtain information on the different $B^0 \to \eta^{(\prime)} \eta^{(\prime)}$
branching ratios and CP asymmetries.

The structure of the amplitudes of the two measured charged decays $B^+ \to K^+ \eta$ and
$B^+ \to K^+ \eta^\prime$ shows that a full cancellation of the dependence on the singlet penguin
contribution ($s^\prime$) is not possible if only these two channels are combined.
In order to cancel completely the residual dependence on these penguins,
information from the corresponding $\Delta S=0$ channels
($B^+ \to \pi^+ \eta$ and $B^+ \to \pi^+ \eta^\prime$) is required.

Since both $|\Delta S|=1$ and $\Delta S=0$ transition amplitudes will be used, it is convenient to make
explicit the different CKM dependence of these amplitudes, as in Eqs.~(\ref{tctcprime})--(\ref{ssprime}).
Notice that at this point we are restricting the validity of the sum rules to the SM
(or to extensions of the SM with the same CKM structure).

In deriving the sum rules that mix $|\Delta S|=1$ and $\Delta S=0$ amplitudes,
we perform the following simplifications.
First, we neglect the contributions of the color-suppressed electroweak penguin amplitude,
$P_{EW}^{C}$, since they are expected to be suppressed \cite{Chiang:2006ih}.
Second, we also neglect the contribution of the color-favored electroweak penguin amplitude,
$P_{EW}$, in the $c$ combination.
Numerically, it is shown to be less than 5\% (in amplitude) of the $C$ contribution
\cite{Chiang:2006ih}.
Third, we redefine the $s^\prime$ combination as
$s^\prime=-\left(Y^u_{sb}+Y^c_{sb}\right)\xi_S\left(S-\frac{1}{3}P_{EW}\right)$
and its contribution is included in all the observables.
The error made by this redefinition is of the order of 1\%.
Finally, we keep the $P_{EW}$ in the $c^\prime$ definition.

The combination of  $|\Delta S|=1$ and $\Delta S=0$ transitions leads to the following
set of four sum rules\footnote{Other sum rules can be found in Ref.~\cite{su3fv2}.}
that we discuss in turn:
\begin{eqnarray} \label{sr1}
{\rm I)} &&
\frac{R^{K\eta}_c}{1-\sqrt{2}\tan\theta}+\frac{R^{K\eta^\prime}_c}{1+\sqrt{2}\cot\theta}
+|r_2|^{2}
\left(\frac{R^{\pi\eta}_0}{1+\tan\theta/\sqrt{2}}+\frac{R^{\pi\eta^\prime}_0}{1-\cot\theta/\sqrt{2}}\right)\cr
&&\cr
&&-\frac{1}{6}\left(4- R_c  + 4 r_1^2 R_c^{\pi\pi}  \right)=q_1\ ,
\end{eqnarray}
where $r_1$ and $r_2$ are defined as
\begin{eqnarray}
r_1&\equiv&\frac{Y^u_{sb}}{Y^u_{db}}=\frac{|V_{us}|}{|V_{ud}|}=0.2318\pm 0.0022\ ,\\[1ex]
|r_2|&\equiv&\left|\frac{Y^u_{sb}+Y^c_{sb}}{Y^u_{db}+Y^c_{db}}\right|=
\left|\frac{|V_{us}||V_{ub}| e^{i \gamma}+|V_{cs}||V_{cb}|}{|V_{ud}||V_{ub}| e^{i\gamma} - |V_{cd}||V_{cb}|}
\right|=4.90^{+0.68}_{-0.56}\ .
\label{r2}
\end{eqnarray}
The numerical values of the CKM elements are taken from Ref.~\cite{Yao:2006px}, and for the angle $\gamma$ we choose the CKMfitter determination
$\gamma=\left(59.0^{+9.2}_{-3.7}\right)^\circ$ \cite{ckmfit}, since it, basically, contains the UTfit determination
$\gamma=\left(64.6\pm4.2\right)^\circ$ \cite{utfit}. Notice that all the dependence on $\gamma$ in the sum rules comes from $|r_2|$. This first sum
rule is the only one in this Letter that can be fully evaluated at present, since all data is available. In the limit of exact flavor $SU(3)$
symmetry $(\xi_T=\xi_C=\xi_P=\xi_S=1)$ $q_1$ is zero (up to the electroweak corrections discussed below), \textit{i.e.}~the experimental value
$q_1^{\rm exp}$ must be compared with zero in this limit. From Table \ref{tableexpvalues}, one obtains
\begin{equation}
\label{q1exp}
q_1^{\rm exp}=0.11\pm 0.65^{+0.19}_{-0.08}\ ,
\end{equation}
where the first error is associated to the branching ratios in Table \ref{tableexpvalues}
and the asymmetric second error comes from the error in $\gamma$.
Interestingly, this sum rule is compatible with zero at the 1$\sigma$ level already in the $SU(3)$ limit.

The inclusion of $SU(3)$ breaking requires taking $\xi_{T,C,P,S}$
in Eqs.~(\ref{tctcprime})--(\ref{ssprime}) different from one.
A scenario such that $\xi_T=\xi_C\equiv\xi_{TC}$ and $\xi_P=\xi_S\equiv\xi_{PS}$
arises in phenomenological analyses, where $\xi_{TC}=f_K/f_\pi$ \cite{Gronau:1994rj} and
$\xi_{PS}=1$ or $\xi_{TC}=\xi f_K/f_\pi$ and $\xi_{PS}=\xi$,
with $\xi$ an universal $SU(3)$-breaking factor.
Here we choose $\xi_{TC}=f_K/f_\pi$ and $\xi_{PS}=\xi$.
The $SU(3)$-breaking parameter $\xi_{TC}$ may be affected by non-factorizable corrections.
However, it was found in Ref.~\cite{Chiang:2006ih} that the best fit to experimental data seems to point out that they are not sizable.
On the contrary, the parameter $\xi$ associated to the penguins accounts for both factorizable and non-factorizable corrections.
In our case, the $SU(3)$-breaking correction to sum rule I is
\begin{eqnarray}
q_1^{SU(3)_{\rm breaking}}=&+&\frac{2}{3} (\frac{f_K^2}{f_\pi^2}-1) r_1^2 R_c^{\pi\pi}\cr
&-& (\xi^2-1)|r_2|^{2}
\left(\frac{R^{\pi\eta}_0}{1+\tan\theta/\sqrt{2}}+\frac{R^{\pi\eta^\prime}_0}{1-
\cot\theta/\sqrt{2}}\right)\ .\qquad
\label{q1su3}
\end{eqnarray}

There is also a remaining electroweak penguin contribution
(originating from the $c^\prime$ definition),
common to sum rule I and III whose explicit form is
\begin{equation}
\label{qEW}
q^{EW}\simeq\frac{1}{3}
\left(-2r_1\frac{\Re [r_2(T+C)P_{EW}^\ast]}{|P|^2}+\left |\frac{P_{EW}}{P}\right |^2\right)\ .
\end{equation}

\begin{figure}[t]
\begin{center}
\psfrag{xi}{$\xi$}
\psfrag{g}{\hspace{-0.6cm}$\gamma$ (deg)}
\includegraphics[width=0.85\textwidth]{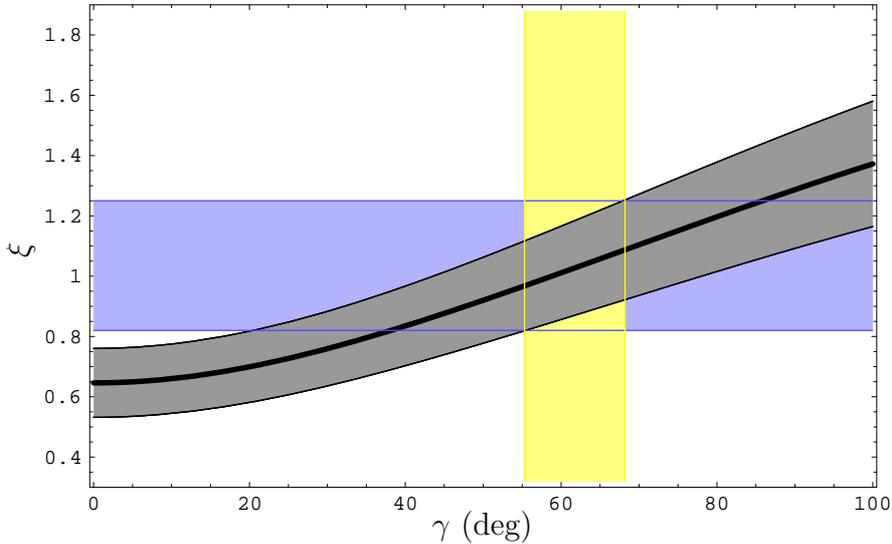}
\end{center}
\caption{\small
The $SU(3)$-breaking parameter $\xi$ as a function of the CKM angle $\gamma$,
as obtained from the ratio $R_{1/0}$ in Eq.~(\ref{ratio10}).
The thick line corresponds to the central values of $R_{1/0}$ and the CKM elements,
and the dark gray band takes into account the experimental errors.
The vertical strip (yellow) corresponds to the SM fit value of gamma
$\gamma_{\rm\sss SM}=(59.0^{+9.2}_{-3.7})^\circ$ \cite{utfit}. The horizontal strip (blue) shows the values
of $\xi$ consistent at 1$\sigma$ with the experimental inputs and $\gamma$.}
\label{plotxigamma}
\end{figure}

A value for $\xi$ can be extracted from the ratio of $|\Delta S|=1$ to $\Delta S=0$ penguin amplitudes
\begin{eqnarray}
\label{ratio10}
R_{1/0}\equiv\frac{\mbox{BR}(B^+\to K^0\pi^+)}{\mbox{BR}(B^0\to K^0\bar K^0)}=
\left|\frac{p^\prime}{p}\right|^2=|r_2|^2\xi^2\ .
\end{eqnarray}
Using the value of $|r_2|$ from Eq.~(\ref{r2}), the experimental result
$R_{1/0}^{\rm exp}=24.1\pm 5.1$ and $\gamma=(59.0^{+9.2}_{-3.7})^\circ$, one gets
\begin{equation}
\xi=1.00\pm 0.15^{+0.08}_{-0.03}\ ,
\label{xi}
\end{equation}
in agreement with recent phenomenological estimates \cite{Chiang:2006ih}.
The dependence of the parameter $\xi$ on the angle $\gamma$ is shown in Fig.~\ref{plotxigamma}.
Notice that if $\xi$ were computed using QCD factorization or any other method,
the ratio $R_{1/0}$ would provide a new way to determine the angle
$\gamma$.

Now we can use the value of $\xi$ in Eq.~(\ref{q1su3}) to estimate the size of the
$SU(3)$ breaking to sum rule I, which gives
\begin{equation}
q_1^{SU(3)_{\rm breaking}}=0.00\pm 0.37\ ,
\end{equation}
and whose error is completely dominated by the error in $\xi$.
The value obtained points to very small $SU(3)$-breaking corrections.
In case this estimation and the present experimental value of sum rule I in Eq.~(\ref{q1exp})
were both confirmed, their difference could be attributed to the electroweak corrections in
Eq.~(\ref{qEW}).

\begin{figure}
\begin{center}
\psfrag{api0eta}{$A_{\rm CP}^{\pi^0\eta}$}
\psfrag{api0etap}
{\begin{minipage}{3cm}\vspace{0.8cm}$A_{\rm CP}^{\pi^0\eta^\prime}$\end{minipage}}
\includegraphics[width=0.85\textwidth]{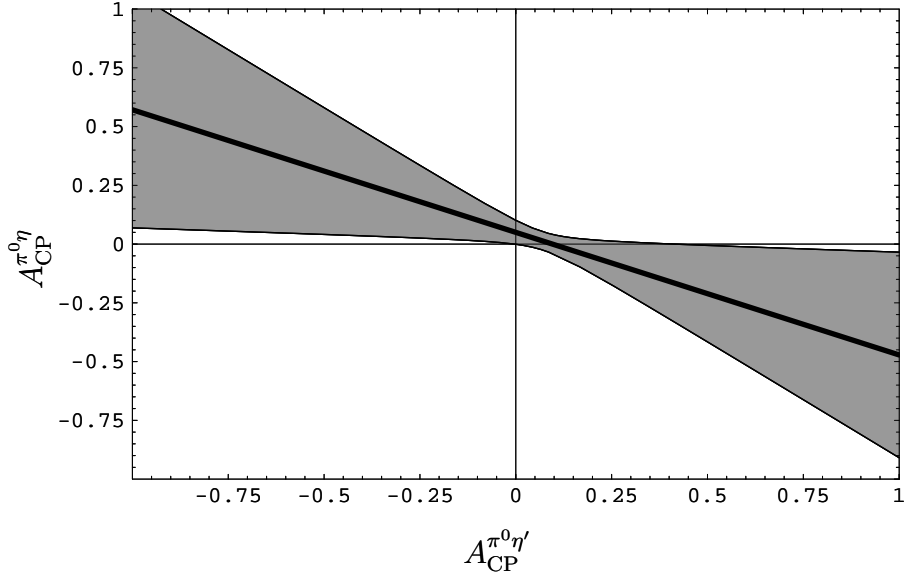}
\end{center}
\caption{\small
$A_{\rm CP}^{\pi^0\eta}$ as a function of $A_{\rm CP}^{\pi^0\eta^\prime}$,
as obtained from sum rule II.
The thick line corresponds to the central values of the observables shown in Table \ref{tableexpvalues}
and the dark gray band takes into account the experimental errors.}
\label{plotAcppi0etaAcppi0etap}
\end{figure}

\begin{eqnarray}  \label{equsumii}
{\rm II)} &&
\frac{R^{K\eta}_c{\cal A}_{CP}^{K^+\eta}}{1-\sqrt{2}\tan\theta}+
\frac{R^{K\eta^\prime}_c{\cal A}_{CP}^{K^+\eta^\prime}}{1+\sqrt{2}\cot\theta}
+|r_2|^{2}
\left(\frac{R^{\pi\eta}_0{\cal A}_{\rm CP}^{\pi^0\eta}}{1+\tan\theta/\sqrt{2}}+
\frac{R^{\pi\eta^\prime}_0{\cal A}_{\rm CP}^{\pi^0\eta^\prime}}{1-\cot\theta/\sqrt{2}}\right)\cr
&&-\frac{1}{6}\left(4{\cal A}_{CP}^{0+}-R_c{\cal A}_{CP}^{+0}+
4r_1^2 R_c^{\pi\pi}{\cal A}_{\rm CP}^{\pi^+\pi^0} \right)=q_2\ .
\end{eqnarray}
This second sum rule is the CP asymmetry partner of the previous one.
In this case we have two unknowns: the two neutral CP asymmetries
${\cal A}_{\rm CP}^{\pi^0\eta}$ and ${\cal A}_{\rm CP}^{\pi^0\eta^\prime}$.
The $SU(3)$-breaking contribution to the sum rule is
\begin{eqnarray} \label{su3q2}
q_2^{SU(3)_{\rm breaking}}= &+&\frac{2}{3}
(\frac{f_K^2}{f_\pi^2}-1)r_1^2 R_c^{\pi\pi}{\cal A}_{\rm
CP}^{\pi^+\pi^0}  \cr &-&
(\xi^2-1)|r_2|^{2}\left(\frac{R^{\pi\eta}_0{\cal A}_{\rm
CP}^{\pi^0\eta}}{1+\tan\theta/\sqrt{2}}+
\frac{R^{\pi\eta^\prime}_0{\cal A}_{\rm
CP}^{\pi^0\eta^\prime}}{1-\cot\theta/\sqrt{2}}\right)\ .\qquad
\end{eqnarray}
If we now use the full sum rule II ---including the $SU(3)$-breaking terms---
taking for $\xi$ the range obtained in Eq.~(\ref{xi}),
we obtain a correlation between the two not yet measured CP asymmetries
${\cal A}_{\rm CP}^{\pi^0\eta}$ and ${\cal A}_{\rm CP}^{\pi^0\eta^{\prime}}$
(see Fig.~\ref{plotAcppi0etaAcppi0etap}).
We observe that for small values of the asymmetries the constrain becomes tight.
Moreover, it predicts a  perfect anticorrelation between both asymmetries for negative values of
${\cal A}_{\rm CP}^{\pi^0\eta^{\prime}}$.
It is worth mentioning that this result is quite robust and very insensitive to $\xi$.
Interestingly, a deviation from this prediction would signal a large electroweak penguin
contribution whose size is expected to be here of order $q^{EW} {\cal A}_{\rm CP}^{\pi^+\pi^0}$.

The next two sum rules to be discussed involve the branching ratio and the CP asymmetry of the
$B^0\to K^0\eta$ decay mode.
The first one contains only the branching ratio:
\begin{equation}
\label{sumrulesKpiexact}
{\rm III)}\quad\frac{R_c^{K\eta}-R_0^{K\eta}}{1-\sqrt{2}\tan\theta}+
\frac{R_c^{K\eta^\prime}-R_0^{K\eta^\prime}}{1+
\sqrt{2}\cot\theta}+\frac{1}{6}(R_0+R_c-4 r_1^2 R_c^{\pi\pi}-2)=q_3\ .\quad
\end{equation}
This sum rule allows to extract the  value of the $B^0 \to K^0 \eta$ branching ratio.
If we now include the $SU(3)$ breaking in the same way as it was done for sum rule I, one finds
\begin{equation}
q_3^{SU(3)_{\rm breaking}}=\frac{2}{3} (\frac{f_K^2}{f_\pi^2}-1)r_1^2 R_c^{\pi\pi}\ .
\label{q3}
\end{equation}
Numerically, the size of the $SU(3)$ breaking in this relation is very small,
$q_3^{SU(3)_{\rm breaking}}=0.0043\pm 0.0004$,
due to the strong suppression factor $r_1^2$.
Then, the predicted value for the CP-averaged branching ratio of $B^0 \to K^0 \eta$
from the full sum rule III ---including $SU(3)$ breaking and up to the aforementioned electroweak corrections--- is
\begin{equation}
\mbox{BR}(B^0 \to K^0 \eta)=(0.38\pm 1.37)\times 10^{-6}\ .
\label{BRk0eta}
\end{equation}
The current experimental bounds on this branching ratio are
$\mbox{BR}(B^0\to K^0\eta)<2.9\cdot 10^{-6}$ (BABAR \cite{exp12})
and $\mbox{BR}(B^0\to K^0\eta)<1.9\cdot 10^{-6}$ (HFAG \cite{HFAG}) at 90\% CL.

Also here there is a remaining contribution from electroweak penguins, $q^{EW}$,
exactly the same as in sum rule I.
As a consequence, the difference between sum rule I and sum rule III
allow us to probe directly the second contribution to $SU(3)$ breaking in Eq.~(\ref{q1su3}).

The last sum rule closing the set of four sum rules that combine $\Delta S=1$ and $\Delta S=0$ transitions is the CP asymmetry partner of sum rule III:
\begin{eqnarray}
{\rm IV)}&&
\frac{{\cal A}_{\rm CP}^{K^+\eta}R_c^{K\eta}-{\cal A}_{\rm CP}^{K^0\eta}R_0^{K\eta}}{1-\sqrt{2}\tan\theta}
            +\frac{{\cal A}_{\rm CP}^{K^+\eta^\prime}R_c^{K\eta^\prime}-
            {\cal A}_{\rm CP}^{K^0\eta^\prime}R_0^{K\eta^\prime}}{1+\sqrt{2}\cot\theta}\cr
&&\cr
&&+\frac{1}{6}({\cal A}_{\rm CP}^{00}R_0+ {\cal A}_{\rm CP}^{+0}R_c -4 r_1^2 R_c^{\pi\pi} {\cal A}_{\rm CP}^{\pi^+\pi^0}
-2 {\cal A}_{\rm CP}^{0+}) = q_4\ . \hspace{2.8cm}
\end{eqnarray}
Again one can obtain the corresponding $SU(3)$-breaking factor,
\begin{equation}
q_4^{SU(3)_{\rm breaking}}=
\frac{2}{3} (\frac{f_K^2}{f_\pi^2}-1)r_1^2 R_c^{\pi\pi}{\cal A}_{\rm CP}^{\pi^+\pi^0}\ ,
\label{q4}
\end{equation}
which is found to be $q_4^{SU(3)_{\rm breaking}}=0.00017\pm 0.00022$ together with a small electroweak penguin contribution of order $q^{EW} {\cal A}_{\rm CP}^{\pi^+\pi^0}$.
Combining the branching ratio obtained from sum rule III with sum rule IV one can obtain a prediction for ${\cal A}_{\rm CP}^{K^0 \eta}$.
However, with the present data the error associated is too large to extract any definite number.

Finally, it is worth noticing that once a measurement for the $\mbox{BR}(B^0 \to K^0 \eta)$ is available
a combination of sum rules III and IV produces a new sum rule for the ${\cal A}_{\rm CP}^{K^0 \eta}$ which is completely independent on the assumed form of the $SU(3)$ breaking and electroweak
corrections,
\begin{eqnarray}
&&\frac{({\cal A}_{\rm CP}^{K^+\eta}-{\cal A}_{\rm CP}^{\pi^+\pi^0})R_c^{K\eta}-
              ({\cal A}_{\rm CP}^{K^0\eta}-{\cal A}_{\rm CP}^{\pi^+\pi^0})R_0^{K\eta}}{1-\sqrt{2}\tan\theta}\cr
&&+\frac{({\cal A}_{\rm CP}^{K^+\eta^\prime}-{\cal A}_{\rm CP}^{\pi^+\pi^0})R_c^{K\eta^\prime}-
                ({\cal A}_{\rm CP}^{K^0\eta^\prime}-{\cal A}_{\rm CP}^{\pi^+\pi^0})R_0^{K\eta^\prime}}
                {1+\sqrt{2}\cot\theta}\\
&&+\frac{1}{6}[({\cal A}_{\rm CP}^{00}-{\cal A}_{\rm CP}^{\pi^+\pi^0})R_0+
                          ({\cal A}_{\rm CP}^{+0}-{\cal A}_{\rm CP}^{\pi^+\pi^0})R_c-
                          2({\cal A}_{\rm CP}^{0+}-{\cal A}_{\rm CP}^{\pi^+\pi^0})]=0\ .\qquad\nonumber
\end{eqnarray}

The next couple of sum rules are closed in the sense that they only involve $\Delta S=0$ processes.
This implies, in particular, that no information on the type of $SU(3)$-breaking corrections is needed and that all electroweak
penguin contributions are included.
\begin{eqnarray}
\label{sr5}
{\rm V)}&&
\frac{B_{\pi^0\eta}/B_{K^0\bar K^0}}{1-\sqrt{2}\tan\theta}+
\frac{B_{\pi^0\eta^\prime}/B_{K^0\bar K^0}}{1+\sqrt{2}\cot\theta}-
\frac{1}{3}-\frac{3}{1-2\sqrt{2}\cot(2\theta)}\\
&&\hspace{-1cm}
\times
\left(\frac{B_{\eta\eta^\prime}}{B_{K^0\bar K^0}}+
\frac{c_{2\theta}+2\sqrt{2}s_{2\theta}+3}{c_{2\theta}+2\sqrt{2}s_{2\theta}-3}
\frac{B_{\eta\eta}}{B_{K^0\bar K^0}}+
\frac{c_{2\theta}+2\sqrt{2}s_{2\theta}-3}{c_{2\theta}+2\sqrt{2}s_{2\theta}+3}
\frac{B_{\eta^\prime\eta^\prime}}{B_{K^0\bar K^0}}
\right)=0\ ,\hspace{0.5cm}\nonumber
\end{eqnarray}
where $B_{\pi^0\eta}\equiv\mbox{BR}(B^0\to \pi^0\eta)$,
$B_{K^0\bar K^0}\equiv\mbox{BR}(B^0\to K^0\bar{K^0})$, etc.
Using the experimental measurements quoted in Table \ref{tableexpvalues},
one obtains for the sum rule the value $0.73^{+0.80}_{-0.63}$,
a result compatible with zero at the 2$\sigma$ level.

The corresponding sum rule for the CP asymmetries
\begin{eqnarray}
\label{sr6}
{\rm VI)}&&
\frac{B_{\pi^0\eta}}{B_{K^0\bar K^0}}\frac{{\cal A}_{\rm CP}^{\pi^0\eta}}{1-\sqrt{2}\tan\theta}+
\frac{B_{\pi^0\eta^\prime}}{B_{K^0\bar K^0}}
\frac{{\cal A}_{\rm CP}^{\pi^0\eta^\prime}}{1+\sqrt{2}\cot\theta}-
\frac{1}{3}{\cal A}_{\rm CP}^{K^0\bar K^0}\nonumber \\
&&
-\frac{3}{1-2\sqrt{2}\cot(2\theta)}\Bigg(
\frac{B_{\eta\eta^\prime}}{B_{K^0\bar K^0}}{\cal A}_{\rm CP}^{\eta\eta^\prime}+
\frac{c_{2\theta}+2\sqrt{2}s_{2\theta}+3}{c_{2\theta}+2\sqrt{2}s_{2\theta}-3}
\frac{B_{\eta\eta}}{B_{K^0\bar K^0}}{\cal A}_{\rm CP}^{\eta\eta}\hspace{1.1cm}\\
&&
+\frac{c_{2\theta}+2\sqrt{2}s_{2\theta}-3}{c_{2\theta}+2\sqrt{2}s_{2\theta}+3}
\frac{B_{\eta^\prime\eta^\prime}}{B_{K^0\bar K^0}}{\cal A}_{\rm CP}^{\eta^\prime\eta^\prime}
\Bigg)=0\ ,\nonumber
\end{eqnarray}
provides a constraint for the specific combination of
${\cal A}_{\rm CP}^{\eta\eta}$, ${\cal A}_{\rm CP}^{\eta^\prime\eta}$ and
${\cal A}_{\rm CP}^{\eta^\prime\eta^\prime}$ entering the sum rule.

\section{Future Prospects}

In this section, we identify which observables have the largest impact on the error size affecting some of the sum rules.
This can be useful as a guide for experimentalists to see which processes could be more interesting to focus on.

\begin{figure}
\begin{center}
\psfrag{g}{\hspace{-0.6cm}$\gamma$ (deg)} \psfrag{xi}{$\xi$}
\includegraphics[width=11cm]{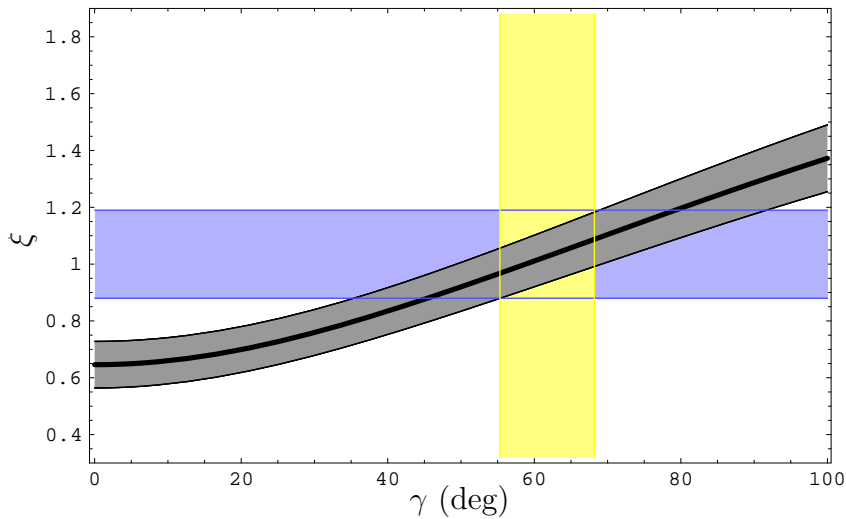}
\end{center}
\caption{\small
The same as Fig.~\ref{plotxigamma} but with the uncertainties of
$\mbox{BR}(B\to K^0 \bar{K^0})$ and $V_{cs}$ reduced by 50\%.
The outcome is a 30\% reduction in the error of $\xi$.}
\label{plot4}
\end{figure}

The extraction of $\xi$ from $R_{1/0}$ in Eq.~(\ref{xi}) is affected mainly by the uncertainties in the branching ratio
$\mbox{BR}(B\to K^0 \bar{K^0})$ and in $V_{cs}$. Fig.~\ref{plot4} shows the impact of decreasing these error bars by a 50\% in the determination
of $\xi$. The result is a 30\% reduction of the error. The error in $\mbox{BR}(B\to K^0 \bar{K^0})$ will be reduced with more
statistics, and the uncertainty in $V_{cs}$, which comes mainly from the error in the lattice determination of the $D\to K$
form factor \cite{Vcs}, is likely to be reduced considerably in future simulations.

\begin{figure}
\begin{center}
\psfrag{api0eta}{$A_{\rm CP}^{\pi^0\eta}$}
\psfrag{api0etap}
{\begin{minipage}{3cm}\vspace{0.8cm}$A_{\rm CP}^{\pi^0\eta^\prime}$\end{minipage}}
\includegraphics[width=11cm]{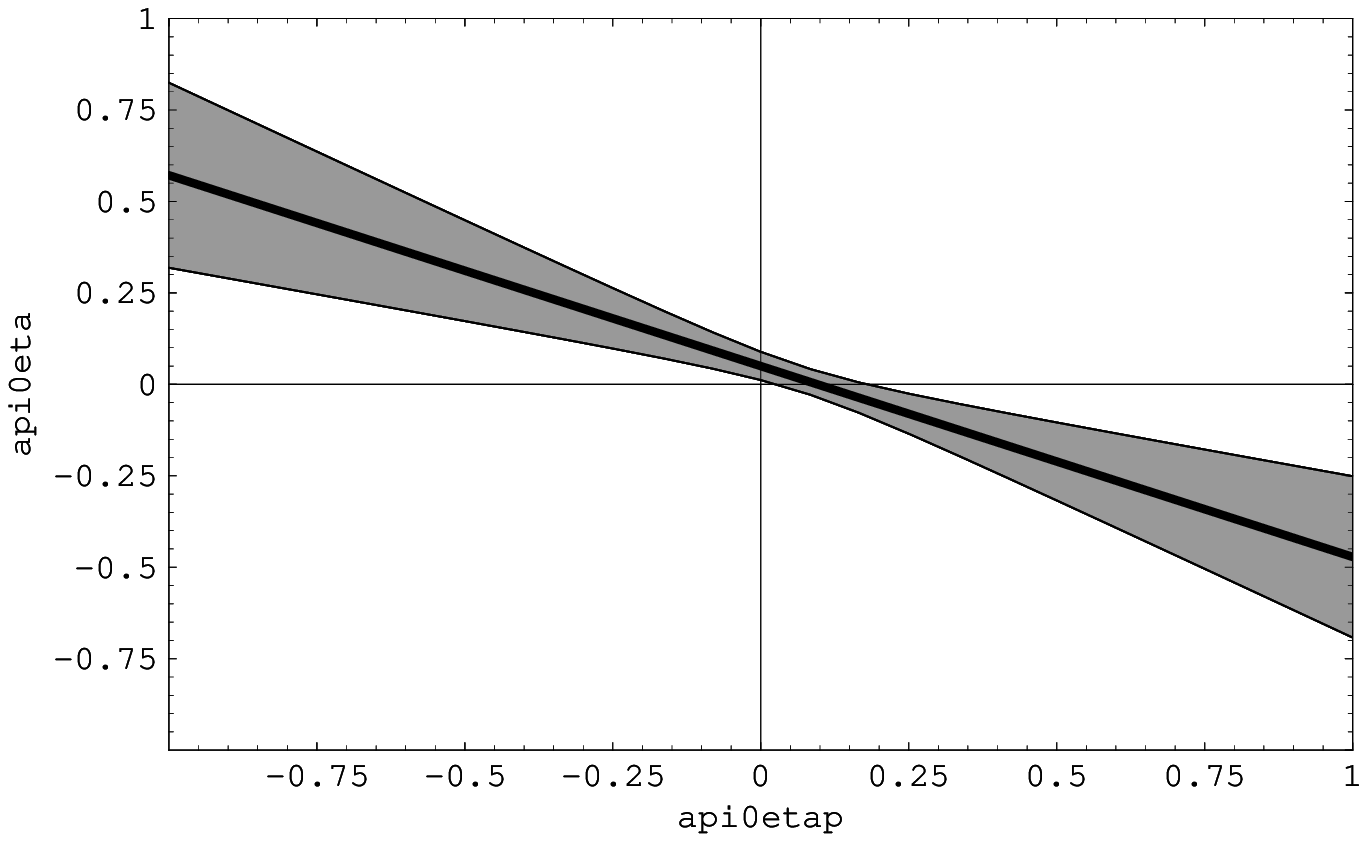}
\end{center}
\caption{\small
The same as in Fig.~\ref{plotAcppi0etaAcppi0etap} but with the uncertainties of $BR(B^0\to \pi^0\eta)$
and $BR(B^0\to \pi^0\eta^\prime)$ reduced by 50\%.}
\label{plot5}
\end{figure}

The situation concerning the CP asymmetries of the two $B \to \pi\eta^{(\prime)}$ decays is depicted in Fig.~\ref{plot5}.
The most important source of uncertainty is due to the branching ratios $\mbox{BR}(B^0\to \pi^0\eta)$
and $\mbox{BR}(B^0\to \pi^0\eta^\prime)$.
Fig.~\ref{plot5} shows the situation in which the uncertainties in these branching ratios
are reduced by a 50\%.
The conclusion is that the predictions for the $B^0\to \pi^0\eta$ CP asymmetry would be up to 50\%
more precise for large values of $A_{\rm CP}^{\pi^0\eta^\prime}$.

Concerning the prediction for the branching ratio $\mbox{BR}(B\to K^0\eta)$, the observables that introduce the dominant
uncertainty are $B^+\to K^+\eta'$ and $B^0\to K^0\eta'$. A 50\% reduction of their error would imply a 35\% reduction
on the uncertainty in the determination of $\mbox{BR}(B\to K^0\eta)$. Of course, these branching ratios are already well measured,
and whether the uncertainties can be reduced by a 50\% is difficult to say.

Therefore, we point out that it would be of utmost importance to focus experimentally on the
$B^0\to \pi^0\eta^{(\prime)}$ modes, specially on the branching ratios,
where a considerable reduction of the uncertainties is experimentally feasible.

\section{Conclusions}
\label{summary}

We conclude with a summary of the main points of this Letter.
We have proposed a series of sum rules based on $B$ decays with $\eta$ and $\eta^\prime$ mesons in the final state.
These sum rules are valid within SM, and include a generic $SU(3)$ breaking scenario.
These $SU(3)$-breaking terms estimate the expected deviation of the sum rules from zero that could be accounted by the SM.
A clear deviation from the numbers given above would be an interesting indication of possible
New Physics  contributions, and would require the revision of the approximations made in deriving the sum rules, mainly the specific contributions of electroweak penguin amplitudes and the choice of the $SU(3)$-breaking scheme.
In Section 3, we have explained how the different contributions from $SU(3)$-breaking or electroweak penguins can be disentangled.
Those sum rules would be sensitive to a large isospin (including electroweak penguins) or
$SU(3)$-breaking New Physics scenario.

The first sum rule, Eq.~(\ref{sr1}), can be already used as a test of the SM, with the $SU(3)$-breaking parameter $\xi$ extracted from $B\to K^0\bar{K}^0$ in Eq.~(\ref{xi}).
If the errors get notably reduced and $\xi$ is obtained from theory,
this sum rule could eventually lead to a determination of the CKM angle $\gamma$.

The second sum rule, Eq.~(\ref{equsumii}), allows to establish correlations between the CP asymmetries of $B^0\to \pi^0\eta$ and $B^0\to \pi^0\eta^\prime$ (Fig.~\ref{plotAcppi0etaAcppi0etap}).
The third sum rule, Eq.~(\ref{sumrulesKpiexact}),
is used to predict the branching ratio $\mbox{BR}(B^0\to K^0\eta)$, see Eq.~(\ref{BRk0eta}).
This prediction can be used in the fourth sum rule to predict the CP asymmetry
$\mathcal{A}_{\rm CP}(B^0\to K^0\eta)$,
but with present data the errors are too big and the result is inconclusive.
We have also provided two sum rules, Eqs.~(\ref{sr5}) and (\ref{sr6}), that involve only $\Delta S=0$ decays, are unaffected by the size of the considered $SU(3)$ breaking and include all the electroweak penguin contributions.

We have pointed out that a reduction in the experimental uncertainties in the branching ratios
$\mbox{BR}(B^0\to \pi^0\eta^{(\prime)})$ and $\mbox{BR}(B\to K\eta^{\prime})$ would suffice to reduce considerably the uncertainties of the predictions for the CP-averaged branching ratio of
$B^0 \to K^0 \eta$, the corresponding CP asymmetry and the error of the combined correlation
${\cal A}_{\rm CP}^{\pi^0 \eta}$--${\cal A}_{\rm CP}^{\pi^0 \eta^\prime}$.

\section*{Acknowledgements}
R.E. gratefully acknowledges the warm hospitality at CERN Theory Division.
J.M. acknowledges long night discussions with Jana Matias.
This work is partially funded by the Ramon y Cajal program (R.E.~and J.M.),
the UAB program PNL2005-41 (J.M.),
the Ministerio de Educaci\'on y Ciencia under grant FPA2005-02211,
the EU, MRTN-CT-2006-035482, FLAVIANET network, and
the Generalitat de Catalunya under grant 2005-SGR-00994.

\end{document}